\documentclass[twocolumn,a4paper]{article}
\usepackage{verbatim}
\usepackage{amssymb}
\usepackage{amsmath}
\usepackage{amsthm}
\usepackage{graphicx}
\usepackage{hyperref}
\usepackage{authblk}

\title{Energy-momentum conservation and Lipkin's zilch}
\author[1]{H. Lashkari-Ghouchani\thanks{hadilashkari@gmail.com}}
\author[2,3]{M.H. Alizadeh\thanks{halizade@bu.edu}}
\affil[1]{\textit{Faculty of Science, PNUM. Mashhad Iran.}}
\affil[2]{\textit{Physics Department, Boston University, Boston, MA 02215}}
\affil[3]{\textit{Photonic Center, Boston University, Boston, MA 02215}}

\begin{document}
\maketitle
\makeatletter
\def\blfootnote{\xdef\@thefnmark{}\@footnotetext}
\makeatother
\blfootnote{Corresponiding author: hadilashkari@gmail.com}
\begin{abstract}
As Noether's theorem states any differentiable symmetry of the action of a physical system has a corresponding conservation law. Lipkin introduced the conservation laws of zilches. But the corresponding symmetries are yet to be determined. Here we find a method to determine those symmetries and by direct calculations express the zilch tensor's relation to current-density for $n$-dimensional Minkowski space-time. Also, we extend this method to calculate symmetries of conservation of energy-momentum. Finally we show that in the special case of n=4, our general equation governing optical chirality reduces to the continuity equation of optical chirality density and optical chirality flow. 
\end{abstract}
\section*{INTRODUCTION}
As Lipkin proved, there are ten conserved quantities for four dimensional electromagnetic field in vacuum which are shown to be independent of energy-momentum\cite{lipkin}. He clarified that the new conserved quantities possess different flow properties compared to energy-momentum\cite{lipkin}. Originally, Lipkin called those conserved quantities \textit{zilch}, the most renown of which is $00$-zilch, $z^{00}$, the \textit{optical chirality}. However he did not find the responsible symmetries for the conservation of the zilches. Recently there has been a growing interest in gaining further insight into these symmetries. For instance, an approach to discuss these symmetries and their connection with helicity's symmetry is presented in \cite{cameron}. In \cite{philbin} Philbin finds the symmetries that are restricted on optical chirality. Also, Bliokh and Nori clarified the connection between helicity, optical chirality and energy which explains why we will find a symmetry of zilches similar to that of helicity\cite{bliokh1,cameron,heaviside,larmor}.\\
Here we find the symmetries of all zilches' conservations, and by applying them into the action of the standard electromagnetic fields, $\frac{1}{16\pi} F_{\alpha\beta}F^{\alpha\beta}$, derive the conservation laws of zilches and their connection with current-density. In doing so and in order to simplify our calculations we use linear form notation. This dramatically reduces our dependence on indices. Then we generalize these symmetries and relations to n-dimensional Minkowski space-time. In the first step, we derive energy-momentum conservation by linear forms and then we apply the same method to the symmetries of zilches to prove their conservation laws. Also we demonstrate the advantage of our approach in finding energy-momentum conservation. Unlike other approaches, where one has to enter the gauge invariance of the energy-momentum tensor manually in to the theory\cite{bliokh,belinfant}, our approach automatically derives this gauge invariance.  \\

\section*{LAGRANGIAN $n$-form}
As a reminder of linear form machinery, here we derive Maxwell's equations from standard electromagnetic Lagrangian by linear forms. Consider an $n$-dimensional Minkowski space-time with $x^\mu$ coordinates. In general we can write an m-form by
$$
a^m=a_{\mu_1\mu_2\ldots\mu_m}dx^{\mu_1}\wedge dx^{\mu_2}\wedge \ldots \wedge dx^{\mu_m}
$$
or summarizing by
\begin{align}\label{formdef}
a^m=a_{\mu^m}dx^{\mu^m}
\end{align}
where $\wedge$ is the \textit{exterior product} or \textit{wedge product}, which makes a completely antisymmetric tensor\cite{frankel}. One must be careful about the upper and lower index of indices. The lower index of indices denotes the number of that index, for example $\mu_m$ is the $m$th index, while the upper index of indices indicates the counting indices of a set, for instance, $\mu^m=\mu_1\mu_2\ldots\mu_m$. If we carefully look at \eqref{formdef} we will notice that if $a_{\mu^m}$ is a complete antisymmetric tensor then every term in \eqref{formdef} is repeated $m!$ times, where $!$ is the factorial. To avoid this repetition we use an arrow, $a_{\underrightarrow{\mu^m}}$, to assume $\mu_1<\mu_2<\ldots<\mu_m$ such that
$$
a^m=\frac{1}{m!}a_{\mu^m}dx^{\mu^m}=a_{\underrightarrow{\mu^m}}dx^{\mu^m}.
$$
Also the exterior differentiation is defined as:
$$
da^m:=da_{\underrightarrow{\mu^m}}\wedge dx^{\mu^m}=\partial_\alpha a_{\underrightarrow{\mu^m}} dx^\alpha\wedge dx^{\mu^m}.
$$
So by $n$-potential $A=A_\beta dx^\beta$ we define the electromagnetic 2-form as follows:
\begin{align}\label{fielddef}
F&=F_{\underrightarrow{\mu^2}} dx^{\mu^2}=\frac{1}{2}F_{\mu^2} dx^{\mu^2}\nonumber\\
&=dA=dA_\beta\wedge dx^\beta=\partial_\alpha A_\beta dx^\alpha\wedge dx^\beta.
\end{align}
which means $F_{\alpha\beta}=\partial_\alpha A_\beta-\partial_\beta A_\alpha$. The immediate consequence of this definition is
\begin{align}\label{max1}
dF=0.
\end{align}
To write Lagrangian and evaluate its variation, we need Hodge operator, $*$, \cite{frankel}. We can define dual of an $m$-form, $a^m$, which is a $(n-m)$-form, $*a^m$, by
\begin{flalign}
*a^m
:=&a^*_{\underrightarrow{\mu^{n-m}}}dx^{\mu^{n-m}}
=\sqrt{|g|} a^{\nu^m}\epsilon_{\underrightarrow{\nu^m}\underrightarrow{\mu^{n-m}}}dx^{\mu^{n-m}}\nonumber\\
=&a_{\alpha^m}g^{\alpha_1\nu_1}g^{\alpha_2\nu_2}\dots g^{\alpha_m\nu_m}\epsilon_{\underrightarrow{\nu^m}\underrightarrow{\mu^{n-m}}}dx^{\mu^{n-m}}
\end{flalign}
where $g$ is the determinant of the metric in $n$-Minkowski space-time, $g=det(g_{\alpha\beta})=-1$, $g^{\alpha\beta}$ is inverse of the metric tensor and $\epsilon_{\beta^n}$ is the Levi-Civita symbol, which is completely antisymmetric and $\epsilon_{123...n}=1$. For the important case of 0-form constant function, $f=1$, we can define $vol^n:=\sqrt{|g|}dx^{12...n}=*1$. We will then have
\begin{align}\label{starreplace}
\langle a^m,b^m \rangle vol^n&:= a_{\underrightarrow{\mu^m}}b^{\mu^m} vol^n\nonumber\\
&=a^m\wedge*b^m = b^m\wedge*a^m.
\end{align}
also one can easily show
\begin{align}\label{starstar}
*(*a^m)=-(-1)^{m(n-m)}a^m.
\end{align}
Now that the necessary definitions are made we can proceed to write the action:
\begin{flalign}
S&:=S_F+S_M=\int_M \mathcal{L}\\
S_F&:=\frac{1}{16\pi}\int_M F_{\alpha\beta}F^{\alpha\beta}vol^n =\frac{1}{8\pi}\int_M F\wedge *F \nonumber\\
S_M&:=-\int_M A_\alpha J^\alpha vol^n=-\int_M A\wedge *J\nonumber
\end{flalign}
which $S_F$ is the field action, $S_M$ is the matter action, $\mathcal{L}$ is the Lagrangian $n$-form and the integrals are defined over the space-time $M$. Using the variation $A\rightarrow A+\delta A$ and  \eqref{fielddef}, the variation of $S_F$ becomes $\delta S_F=\frac{1}{4\pi}\int_M\delta A\wedge d*F+\frac{1}{4\pi}\int_M d(\delta A\wedge *F)$. Also $\delta S_M=-\int_M \delta A\wedge*J$ then
\begin{align}\label{delS}
\delta S&=\int_M \delta A\wedge*\Big(\frac{1}{4\pi}{(-1)}^n*d*F-J\Big)\nonumber\\
&+\int_M d(\delta A\wedge *F)=0
\end{align}
where we have used \eqref{starstar}. The second term in \eqref{delS} is the integral over the boundary of space-time
$$
\int_M d(\delta A\wedge *F)=\int_{\partial M} \delta A\wedge *F
$$
which is zero and based on Noether's theorem $\delta A\wedge *F$ is the flow density tensor of a conserved quantity\cite{jose,weinberg}. So the first term determines the equations of motion which are
\begin{align}\label{eqmotion}
*d*F=4\pi{(-1)}^nJ.
\end{align}
\eqref{max1} and \eqref{eqmotion} are the \textit{Maxwell's equations}. Conservation of electric charge can be found by:
$
*J=\frac{1}{4\pi}d*F
$
so
$
d*J=0
$
which means electric charge that passes through region $U^{n-1}$, $ 4\pi q=\int_U 4\pi *J=\int_U d*F=\int_{\partial U} *F$, is a conserved quantity. It is worth mentioning that we did not use any indices to calculate equations of motions and conservation equations. We will express symmetries of this action and their conserved quantities in this powerful notation. One can expand equations of motion, \eqref{eqmotion}, to find
\begin{align}\label{JF}
\partial_\beta F^{\alpha\beta}=4\pi J^\alpha
\end{align}
detailed derivation can be found in the Appendix.

\section*{ENERGY-MOMENTUM}
Energy-momentum tensor of an electromagnetic field can be written as:
\begin{align}\label{energym}
T_{\alpha\beta}=F_{\alpha\gamma}F^{\ \ \gamma}_{\beta}-\frac{1}{4}g_{\alpha\beta}F_{\mu\nu}F^{\mu\nu}.
\end{align}
In this section we will derive this energy-momentum tensor. These calculations are to demonstrate the advantage of our approach as it does not necessitate the appearance of indices, one does not have to manually insert divergence-free terms and finally the gauge invariance of energy-momentum is automatically fulfilled. we start from the infinitesimal variation of $h_\alpha$ over space-time, which is the transformation
\begin{align}\label{varform}
F\rightarrow F+h^\alpha\partial_\alpha F.
\end{align}
In this step we can have two approaches to make the variation of the Lagrangian $n$-form a closed form. Each of these approaches are coming from two different identities. In the end we will use the results of both approaches to comprehend the conservation laws. The first identity is as follows
\begin{flalign}\label{firidentity}
d*(* F\wedge g_{\beta\lambda}dx^\lambda)
=-\partial_\beta F
\end{flalign}
details of derivation can be found in Appendix. The next one is
\begin{align}\label{secidentity}
*d*(F\wedge dx^\beta)=4\pi J\wedge dx^\beta-\partial_\alpha F g^{\alpha\beta}.
\end{align}
using these identities, the variation of \eqref{varform} for the first approach becomes
\begin{align}\label{gvarformfirst}
F\rightarrow F-h_\alpha d*(*F\wedge dx^\alpha)
\end{align}
and for the second approach becomes
\begin{align}\label{gvarformsecond}
F\rightarrow F+h_\alpha\Big(-*d*(F\wedge dx^\alpha)+4\pi J\wedge dx^\alpha\Big).
\end{align}
Considering an infinitesimal variation of the action, using $\delta S_M=0$ and \eqref{varform}, for the first approach we have
\begin{flalign}
\delta S
&=-\frac{1}{4\pi}h_\alpha\int_M d*(*F\wedge dx^\alpha)\wedge*F\nonumber\\
&=-\frac{1}{4\pi}h_\alpha\int_M d\Big(*(*F\wedge dx^\alpha)\wedge*F\Big)\nonumber\\
&\qquad\qquad+4\pi J\wedge dx^\alpha\wedge * F
\end{flalign}
and for the next approach
\begin{flalign}
\delta S
&=\frac{1}{4\pi}h_\alpha\int_M F\wedge * \Big(-*d*(F\wedge dx^\alpha)\nonumber\\
&\qquad\qquad+4\pi J\wedge dx^\alpha\Big)\nonumber\\
&=\frac{1}{4\pi}h_\alpha\int_M d\Big(F\wedge *(F\wedge dx^\alpha)\Big)\nonumber\\
&\qquad\qquad+4\pi J\wedge dx^\alpha\wedge * F
\end{flalign}
where we used \eqref{max1}, \eqref{starreplace}, \eqref{starstar} and \eqref{eqmotion}. Here we can argue if there is no $n$-current-density, $J^\alpha=0$, then both of these integrands become closed forms, hence based on Noether's theorem we can subtract these two closed forms to find $dT^\alpha=0$\cite{jose,weinberg}. Then, we have $n$ conserved quantities, $p^\alpha=\int_U T^\alpha$, that pass through region $U^{n-1}$. These $n$ conserved quantities are the $n$-vector of energy-momentum of the field. In the general case when $n$-current-density, $J$, is non-zero we have
\begin{flalign}\label{emconservation}
dT^\alpha&=-\frac{1}{2}d\Big(*(*F\wedge dx^\alpha)\wedge*F+F\wedge *(F\wedge dx^\alpha)\Big)\nonumber\\
&=4\pi J\wedge dx^\alpha\wedge * F
\end{flalign}
where $T^\alpha$ denotes the \textit{energy-momentum tensor}, the components of which can be shown to to be the same as \eqref{energym}, see the Appendix. The simplicity of the proof and the fact that we did not have to go through Belinfante symmetrization procedure for the energy-momentum tensor add to the power of our method. Some other approaches can be found here\cite{belinfant,bliokh,jose}. Further, the obtained tensor is naturally gauge invariant. We mention in passing that the conservation of angular momentum can be addressed by replacing transformation operator, $h^\alpha \partial_\alpha$, in \eqref{varform} by rotation operator, $ \omega^{\alpha\beta}(x_\alpha\partial_\beta-x_\beta\partial_\alpha)$, so the variation of electromagnetic field for infinitesimal rotation is
\begin{align}
F\rightarrow F+\omega^{\alpha\beta}(x_\alpha\partial_\beta-x_\beta\partial_\alpha) F.\nonumber
\end{align}
Going through the same procedure the angular momentum tensor of the electromagnetic field can be found to be
\begin{flalign}
dM^{\alpha\beta}
&=-\frac{1}{2}d\Big(*(*F\wedge (x^\alpha dx^\beta-x^\beta dx^\alpha))\wedge*F\nonumber\\
&\qquad\qquad+F\wedge *(F\wedge (x^\alpha dx^\beta-x^\beta dx^\alpha))\Big)\nonumber\\
&=4\pi J\wedge (x^\alpha dx^\beta-x^\beta dx^\alpha)\wedge * F.\nonumber
\end{flalign}
which demonstrates the conservation of angular momentum in absence of $n$-vector of current-density, $dM^{\alpha\beta}=0$.

\section*{Zilch}
The generalization of zilch tensor as Lipkin defined\cite{lipkin} to $n$-dimensional Minkowski space-time can be achieved by using the following variation
\begin{align}\label{chvar}
F\rightarrow F+l^{\alpha\beta}*(N^{n-4}\wedge \partial_\alpha \partial_\beta F)
\end{align}
where $N^{n-4}$ is a constant $(n-4)$-form, and $l^{\alpha\beta} $ are infinitesimal parameters of the variation. It is important to note that, because of $N$, which is a $(n-4)$-form, \eqref{chvar} is defined only for $n\geq 4$. This variation is the main result of this paper. For $n=4$, this variation reduce to
\begin{flalign}
&\mathbf{E}\rightarrow \mathbf{E}-l^{\alpha\beta} \partial_\alpha \partial_\beta \mathbf{B}\nonumber\\
&\mathbf{B}\rightarrow \mathbf{B}+l^{\alpha\beta} \partial_\alpha \partial_\beta \mathbf{E}.\nonumber
\end{flalign}
Focusing on optical chirality, for a \textit{monochromatic} field this variation becomes
\begin{flalign}
&\mathbf{E}\rightarrow \mathbf{E}+l^{00} \omega^2 \mathbf{B}=\mathbf{E}+\theta \mathbf{B}\nonumber\\
&\mathbf{B}\rightarrow \mathbf{B}-l^{00} \omega^2 \mathbf{E}=\mathbf{B}-\theta \mathbf{E}\label{helsym}
\end{flalign}
where $\mathbf{E}$ is the electric and $\mathbf{B}$ the magnetic vector in cgs units, $\omega$ is the angular frequency, $\theta$ is an infinitesimal parameter of electric-magnetic rotation and \eqref{helsym} is the infinitesimal form of the \textit{dual symmetry} or \textit{duplex transformation}\cite{bliokh,barnet,cameron}. The dual symmetry is associated with optical helicity so the connection between special case of zilches, optical chirality, with helicity becomes more clear\cite{bliokh1}. To find conserved quantities, again we have two approaches to make the variation of Lagrangian $n$-form a closed form, then we can subtract those two closed forms to find the conservation laws. To get to that point we need two identities. Based on Appendix, the first identity is \eqref{chfiridentity0}
\begin{flalign}
*\Big(N\wedge d*(* \partial_\alpha F\wedge g_{\beta\lambda}dx^\lambda)\Big)
=-*(N\wedge \partial_\alpha\partial_\beta F)\label{chfiridentity}
\end{flalign}
and the second one is
\begin{align}\label{chsecidentity}
g_{\beta\gamma}&d*(N\wedge \partial_\alpha F\wedge dx^\gamma)=*(N\wedge \partial_\alpha \partial_\beta F)\nonumber\\
&+ 4\pi g_{\beta\gamma}*(N\wedge dx^\gamma\wedge \partial_\alpha J)-U_{\alpha\beta}.
\end{align}
Where we have
\begin{align}\label{ugly}
U_{\alpha\beta}:=(n-4)\epsilon_{\underrightarrow{\mu^{n-5}}\underrightarrow{\nu^2}\underrightarrow{\gamma^2}\beta}\partial_\alpha\partial_{\theta} F^{\nu^2} N^{\mu^{n-5}\theta}dx^{\gamma^2}.
\end{align}
This term is of great significance. This is because it neither makes a closed form when substituted into the action, nor is it related to the $n$-current-density. Using \eqref{chfiridentity} and \eqref{chsecidentity} identities the variation of \eqref{chvar} for first approach becomes
\begin{flalign}\label{chvarcomfirst}
F\rightarrow F-l^{\alpha\beta}g_{\beta\gamma}*\Big(N\wedge d*(*\partial_\alpha F\wedge dx^\gamma)\Big)
\end{flalign}
and for the second approach
\begin{flalign}\label{chvarcomsecond}
F\rightarrow F&+l^{\alpha\beta}g_{\beta\gamma}\bigg(d*(N\wedge \partial_\alpha F\wedge dx^\gamma)\nonumber\\
&-4\pi g_{\beta\gamma}*(N\wedge dx^\gamma\wedge \partial_\alpha J)+U_{\alpha\beta}\bigg).\nonumber\\
\end{flalign}
So for the first approach the variation of the action becomes
\begin{flalign}\label{chactionvar}
\delta S=&\frac{1}{4\pi}l^{\alpha\beta} g_{\beta\gamma}\int_M F\wedge N\wedge d*(*\partial_\alpha F\wedge dx^\gamma)
\end{flalign}
and for the second approach it becomes
\begin{flalign}\label{chactionvar}
\delta S=&\frac{1}{4\pi}l^{\alpha\beta} g_{\beta\gamma}\int_M \Big(d*(N\wedge \partial_\alpha F\wedge dx^\gamma)\nonumber\\
&-4\pi *(N\wedge dx^\gamma\wedge \partial_\alpha J)+U_{\alpha}^{\ \gamma}\Big)\wedge *F.
\end{flalign}
Using \eqref{max1}, \eqref{starreplace}, \eqref{starstar}, \eqref{eqmotion} and Noether's theorem, zilch tensor, $Z_{\alpha}^{\ \gamma}$, is found the same way as for energy-momentum tensor.
\begin{flalign}
dZ_{\alpha}^{\ \gamma}&=d\Big(F\wedge *(*\partial_\alpha F\wedge dx^\gamma)\wedge N\nonumber\\
&\quad-*(N\wedge\partial_\alpha F \wedge dx^\gamma)\wedge *F\Big)\label{chconservation}\\
&=4\pi N\wedge\Big(J\wedge \partial_\alpha F-\partial_\alpha J\wedge F\Big)\wedge dx^\gamma\nonumber\\
&\quad+U_{\alpha}^{\ \gamma}\wedge *F\label{currentofch}.
\end{flalign}
This is the second most important result of this paper, which defines \textit{zilch tensor} in $n$-Minkowski space-time and its relation to $n$-current-density. If all terms of \eqref {currentofch} go to zero, then there is a conserved quantity, $z_{\alpha}^{\ \gamma}=\int_U Z_{\alpha}^{\ \gamma}$, that passes through region $U^{n-1}$ which Lipkin named \textit{zilch}. To see why this term is Lipkin's zilch tensor we need to expand it. For the first term of $Z_{\alpha}^{\ \gamma}$ in \eqref{chconservation} this procedure is less cumbersome if we multiply $dx^\phi$ to them
\begin{flalign}
&dx^\phi\wedge F\wedge *(g^{\alpha\gamma}*\partial_\gamma F\wedge dx^\beta)\wedge N\nonumber\\
&=g^{\alpha\gamma}g^{\beta\sigma} \epsilon^{\phi\nu^2\lambda\mu^{n-4}}F_{\underrightarrow{\nu^2}}\partial_\gamma F_{\lambda\sigma}N_{\underrightarrow{\mu^{n-4}}}vol^n.
\end{flalign}
Also, the second term of $Z_{\alpha}^{\ \gamma}$ in \eqref{chconservation} can be extracted as follows
\begin{flalign}
&-dx^\phi \wedge *(N\wedge g^{\alpha\gamma}\partial_\gamma F \wedge dx^\beta)\wedge *F\nonumber\\
&=-g^{\alpha\gamma_1}g^{\beta\gamma_2}\epsilon^{\phi\kappa\theta^{n-2}}\epsilon_{\underrightarrow{\mu^{n-4}}\underrightarrow{\nu^2}\gamma_2\kappa}\epsilon_{\underrightarrow{\sigma^2}\underrightarrow{\theta^{n-2}}}\nonumber\\
&\qquad\qquad \qquad\qquad \qquad\quad\times N^{\mu^{n-4}}\partial_{\gamma_1}(F^{\nu^2})F^{\sigma^2} vol^n\nonumber\\
&=-det(g^{\sigma\theta})g^{\alpha\gamma}g^{\phi\psi}\epsilon^{\overrightarrow{\mu^{n-4}}\overrightarrow{\nu^2}\beta\kappa}N_{\mu^{n-4}}\partial_{\gamma}(F_{\nu^2})F_{\psi\kappa} vol^n.
\end{flalign}
In case of $n=4$ zilch tensor becomes
\begin{flalign}
Z^{\alpha\beta\phi}vol^4&:=dx^\phi \wedge Z^{\alpha\beta}\nonumber\\
&=N g^{\alpha\gamma}\Big(g^{\phi\psi}\epsilon^{\nu^2\beta\kappa} \partial_{\gamma}(F_{\underrightarrow{\nu^2}})F_{\psi\kappa}\nonumber\\
&\qquad\qquad +g^{\beta\sigma} \epsilon^{\phi\nu^2\lambda}F_{\underrightarrow{\nu^2}}\partial_\gamma F_{\lambda\sigma}\Big)vol^4 \nonumber
\end{flalign}
where $N$ is now a number. If $N=1$ this expression is equal to the first four terms of original definition of Lipkin's zilch tensor, equation (4) of original Lipkin's paper\cite[p. 2]{lipkin}. Care must be taken in treating \eqref{chfiridentity} and \eqref{chsecidentity}, which are in 4-Minkowski space-time. This is because in the absence of 4-current-density they are symmetric upon exchange of $\alpha\beta$. Additionally, one can show the four last terms in the original Lipkin's definition\cite[p. 2]{lipkin} make a closed form, therefore, they can be removed from the definition.\\
Let's  expand the last term of \eqref{currentofch}
\begin{flalign}\label{uglycurrent}
&U_{\alpha\beta}\wedge*F\nonumber\\
&\quad=(n-4)N^{\mu^{n-5}\theta}\epsilon_{\underrightarrow{\mu^{n-5}}\underrightarrow{\gamma^2}\underrightarrow{\nu^2}\beta}F^{\gamma^2}\partial_\theta\partial_\alpha F^{\nu^2}vol^n.
\end{flalign}
It is noted that for $n=4$, $U_{\alpha\beta}=0$ and \eqref{uglycurrent} can be disregarded for $n=4$ theories. However for $n>4$ and $J=0$, \eqref{uglycurrent} predicts generation and anihilation of zilches inside of a region, which lacks in $n=4$ theories. \\
As the final remark we derive the governing equation for optical chirality in $n=4$ from the general formulas \eqref{chconservation}-\eqref{currentofch}. One can easily show that in the case of $n=4$ \eqref{chconservation}-\eqref{currentofch} reduces to
\begin{flalign}
&\partial_t\big(\mathbf{B}.\partial_t\mathbf{D}-\mathbf{D}.\partial_t\mathbf{B}\big)+\nabla.\big(\mathbf{E}\times\partial_t\mathbf{D}+\mathbf{H}\times\partial_t\mathbf{B}\big)\nonumber\\
&=4\pi (\mathbf{J}.\partial_t\mathbf{B}-\mathbf{B}.\partial_t\mathbf{J}).\nonumber
\end{flalign}
where in the case of $\mathbf{J}=0$ we get the continuity equation of optical chirality
\begin{flalign}
&\partial_t\big(\mathbf{B}.\nabla\times\mathbf{H}+\mathbf{D}.\nabla\times\mathbf{E}\big)\nonumber\\
&+\nabla.\big(\mathbf{E}\times\nabla\times\mathbf{H}-\mathbf{H}\times\nabla\times\mathbf{E}\big)=0\nonumber
\end{flalign}
where $\mathbf{B}.\nabla\times\mathbf{H}+\mathbf{D}.\nabla\times\mathbf{E}$ is the \textit{optical chirality density} and $\mathbf{E}\times\nabla\times\mathbf{H}-\mathbf{H}\times\nabla\times\mathbf{E}$ is the \textit{optical chirality flow}.

\section*{CONCLUSION}
In this paper we studied the symmetries of standard electromagnetic action. First, we obtained two identities to calculate energy-momentum tensor directly from transformation symmetry without manually inserting any divergence-free terms. Then we found a new symmetry and by applying the same method as for the energy-momentum tensor, we generalized Lipkin's zilch which was originally defined in $n=4$ to $n$-dimensional Minkowski space-time. In our calculations we used linear forms to avoid complexity of indices. In all the important results of this paper the $n$-current-density was kept non-zero.

\section*{APPENDIX: IDENTITIES}
Here we derive and expand in detail the identities referred in the main text.
\begin{flalign}\label{calcexample1}
*d*F&=*d*\left(F_{\underrightarrow{\mu^2}}dx^{\mu^2}\right)=*d\left(F^{\mu^2}\epsilon_{\underrightarrow{\mu^2}\underrightarrow{\nu^{n-2}}}dx^{\nu^{n-2}}\right)\\
&=det(g^{\sigma\theta})\epsilon^{\beta\nu^{n-2}\alpha}\epsilon_{\underrightarrow{\mu^2}\underrightarrow{\nu^{n-2}}}\partial_\beta F^{\mu^2}g_{\alpha\gamma}dx^\gamma\nonumber\\
&={(-1)}^n\frac{1}{2}(\delta^\alpha_{\mu_1}\delta^\beta_{\mu_2}-\delta^\alpha_{\mu_2}\delta^\beta_{\mu_1})\partial_\beta F^{\mu^2}g_{\alpha\gamma}dx^\gamma\nonumber\\
&={(-1)}^n\partial_\beta F^{\alpha\beta}g_{\alpha\gamma}dx^\gamma.\nonumber
\end{flalign}
Based on \eqref{eqmotion}, it must be equal to
$
4\pi{(-1)}^n J_\gamma dx^\gamma
$
so
$$
\partial_\beta F^{\alpha\beta}=4\pi J^\alpha\qquad\qed
$$
In energy-momentum section, we needed two identities. For the first identity, we try to expand $*\Big(N\wedge d*(* F\wedge g_{\beta\lambda}dx^\lambda)\Big)$. To do so we assume a free constant $(n-4)$-form, $N^{n-4}$ and use the following identity
\begin{align}\label{antisymiden}
\epsilon_{[\mu^n}g_{\mu_{n+1}]\beta}=0
\end{align}
which is because a term with $n+1$ completely antisymmetric indices, in $n$-dimensional space-time, is always zero. Now we can write
\begin{flalign}
&*\Big(N\wedge d*(* F\wedge g_{\beta\lambda}dx^\lambda)\Big)\nonumber\\
&\qquad=*\Big(N\wedge d(F_{\nu\beta}dx^\nu)\Big)\nonumber\\
&\qquad=\epsilon_{\underrightarrow{\mu^{n-4}}\lambda\nu_1\underrightarrow{\gamma^2}}g_{\nu_2\beta}\partial^\lambda  F^{\nu^2} N^{\mu^{n-4}}dx^{\gamma^2}\nonumber\\
&\qquad=-\epsilon_{\underrightarrow{\mu^{n-4}}\underrightarrow{\nu^2}\underrightarrow{\gamma^2}}g_{\lambda\beta}\partial^\lambda  F^{\nu^2} N^{\mu^{n-4}}dx^{\gamma^2}\nonumber\\
&\qquad=-*(N\wedge \partial_\beta F)\label{chfiridentity0}
\end{flalign}
where we used $\partial^{[\lambda}F^{\nu^2]}=0$, which is another version of $dF=0$. As long as $N$ is a free $(n-4)$-form we can find the first identity
$$
d*(* F\wedge g_{\beta\lambda}dx^\lambda)
=-\partial_\beta F\qquad\qed
$$
For the second identity, after lines of algebra similar to \eqref{calcexample1} one can show
\begin{align}
&*d*(F\wedge dx^\beta)\nonumber\\
&\quad=-\frac{1}{4}(4\partial_\alpha F^{\alpha\mu} g^{\nu\beta} + 2\partial_\alpha F^{\mu\nu} g^{\alpha\beta})g_{\mu\gamma}g_{\nu\theta}dx^\gamma \wedge dx^\theta\nonumber\\
&\quad=4\pi J\wedge dx^\beta-\partial_\alpha F g^{\alpha\beta}\qquad\qed\nonumber
\end{align}
To prove that componants of \eqref{emconservation} are the same as those of \eqref{energym} we have
\begin{flalign}
&dx^\phi\wedge*(*F\wedge dx^\alpha)\wedge*F\nonumber\\
&\qquad =\epsilon^{\phi\lambda\mu^{n-2}}F_{\lambda\gamma}g^{\gamma\alpha}\epsilon_{\underrightarrow{\nu^2}\underrightarrow{\mu^{n-2}}}F^{\nu^2}vol^n\nonumber\\
&\qquad =-F^{\phi\lambda}F^\alpha_{\ \lambda}vol^n\nonumber
\end{flalign}
and
\begin{align}
&dx^\phi\wedge F\wedge*(F\wedge dx^\alpha)\nonumber\\
&\qquad =\epsilon^{\phi\nu^2\mu^{n-3}}F_{\underrightarrow{\nu^2}}\epsilon_{\underrightarrow{\sigma^2}\lambda\underrightarrow{\mu^{n-3}}}F^{\sigma^2}g^{\lambda\alpha}vol^n\nonumber\\
&\qquad =\frac{1}{4}(2F_{\nu^2}F^{\nu^2}g^{\phi\alpha}+4F_{\nu^2}F^{\phi\nu_1}g^{\nu_2\alpha})vol^n\nonumber\\
&\qquad =\left(\frac{1}{2}F_{\nu^2}F^{\nu^2}g^{\phi\alpha}-F^{\phi\lambda}F^\alpha_{\ \lambda}\right)vol^n\nonumber
\end{align}
so
\begin{flalign}
&T^{\alpha\phi}vol^n=dx^\phi\wedge T^\alpha\nonumber\\
&\qquad=-\frac{1}{2}dx^\phi\wedge\Big(*(*F\wedge dx^\alpha)\wedge*F\nonumber\\
&\qquad\qquad\qquad+F\wedge *(F\wedge dx^\alpha)\Big)\nonumber\\
&\qquad=\left(F^{\phi\lambda}F^\alpha_{\ \lambda}-\frac{1}{4}F_{\nu^2}F^{\nu^2}g^{\phi\alpha}\right)vol^n\qquad\qed\nonumber
\end{flalign}
which is the same as \eqref{energym}.
In the third section, zilch, the second identity can be achieved by implementing a similar identity to \eqref{antisymiden}, $\epsilon_{[\mu^n}\partial_{\mu_{n+1}]}=0$. Applying it to
\begin{flalign}
&g_{\beta\gamma}d*(N\wedge \partial_\alpha F\wedge dx^\gamma)\nonumber\\
&\qquad=\epsilon_{\underrightarrow{\mu^{n-4}}\underrightarrow{\nu^2}\beta\gamma}\partial_\sigma \partial_\alpha F^{\nu^2} N^{\mu^{n-4}}dx^\sigma \wedge dx^\gamma\nonumber
\end{flalign}
we can get
\begin{flalign}
&=\frac{1}{2}\Bigg(\epsilon_{\underrightarrow{\mu^{n-4}}\underrightarrow{\nu^2}\gamma^2}\partial_\beta+2\epsilon_{\underrightarrow{\mu^{n-4}}\beta\underrightarrow{\gamma^2}\nu_1}\partial_{\nu_2}\nonumber\\
&\quad-(n-4)\epsilon_{\underrightarrow{\mu^{n-5}}\underrightarrow{\nu^2}\beta\gamma^2}\partial_{\mu_{n-4}}\Bigg) \partial_\alpha F^{\nu^2} N^{\mu^{n-4}}dx^{\gamma^2}\nonumber\\
&=*(N\wedge \partial_\alpha \partial_\beta F)+ 4\pi g_{\beta\gamma}*(N\wedge dx^\gamma\wedge \partial_\alpha J)\nonumber\\
&\quad-\frac{1}{2}(n-4)\epsilon_{\underrightarrow{\mu^{n-5}}\underrightarrow{\nu^2}\beta\gamma^2}\partial_{\theta}\partial_\alpha F^{\nu^2} N^{\mu^{n-5}\theta}dx^{\gamma^2}.\nonumber
\end{flalign}
Where we again have
$$
U_{\alpha\beta}:=(n-4)\epsilon_{\underrightarrow{\mu^{n-5}}\underrightarrow{\nu^2}\beta\underrightarrow{\gamma^2}}\partial_{\theta}\partial_\alpha F^{\nu^2} N^{\mu^{n-5}\theta}dx^{\gamma^2}.
$$
So the second identity is
\begin{align}
g_{\beta\gamma}&d*(N\wedge \partial_\alpha F\wedge dx^\gamma)=*(N\wedge \partial_\alpha \partial_\beta F)\nonumber\\
&+ 4\pi g_{\beta\gamma}*(N\wedge dx^\gamma\wedge \partial_\alpha J)-U_{\alpha\beta}\qquad\qed
\end{align}

\bibliography{Biblo}

\begin{thebibliography}{10}

\bibitem{lipkin}
D.M. Lipkin.
\newblock Existence of a new conservation law in electromagnetic theory.
\newblock {\em Journal of Mathematical Physics}, May 1964.

\bibitem{cameron}
R.P Cameron~S.M Barnett and A.~M Yao.
\newblock Optical helicity, optical spin and related quantities in
  electromagnetic theory.
\newblock {\em New J. Phys. 14 053050}, May 2012.

\bibitem{philbin}
T.~G. Philbin.
\newblock Lipkin’s conservation law, noether’s theorem, and the relation to
  optical helicity.
\newblock {\em PHYSICAL REVIEW A 87, 043843}, April 2013.

\bibitem{bliokh1}
K.Y. Bliokh and F.~Nori.
\newblock Characterizing optical chirality.
\newblock {\em PHYSICAL REVIEW A 83, 021803(R)}, February 2011.

\bibitem{heaviside}
O.~Heaviside.
\newblock On the forces, stresses and fluxes of energy in the electromagnetic
  field.
\newblock {\em Phil. Trans. R. Soc. A 183 423–80}, 1892.

\bibitem{larmor}
J.~Larmor.
\newblock Dynamical theory of the electric and luminiferous medium iii.
\newblock {\em Phil. Trans. R. Soc. A 190 205–300}, 1897.

\bibitem{bliokh}
K.Y Bliokh~A.Y Bekshaev and F.~Nori.
\newblock Dual electromagnetism: helicity, spin, momentum and angular momentum.
\newblock {\em New Journal of Physics 15}, March 2013.

\bibitem{belinfant}
F.J. Belinfante.
\newblock On the quantum theory of wave fields.
\newblock {\em Physica 7 449}, 1940.

\bibitem{frankel}
T.~Frankel.
\newblock {\em The geometry of physics : an introduction, 2nd ed.}
\newblock Cambridge ; New York : Cambridge University Press, 2004.

\bibitem{jose}
J.~José.
\newblock {\em Classical dynamics : a contemporary approach}.
\newblock Cambridge [England] ; New York : Cambridge University Press, 1998.

\bibitem{weinberg}
S.~Weinberg.
\newblock {\em The Quantum Theory of Fields}.
\newblock Cambridge University Press, Cambridge, UK, 1995.

\bibitem{barnet}
S.M. Barnett~R.P. Cameron and A.M. Yao.
\newblock Duplex symmetry and its relation to the conservation of optical
  helicity.
\newblock {\em PHYSICAL REVIEW A 86, 013845}, July 2012.

\end{thebibliography}
\bibliographystyle{unsrt}

\end{document}